\documentclass[superscriptaddress,twocolumn, showpacs,amsmath,amssymb,aps,prl,floatfix, 10pt]{revtex4-2}
\usepackage[american]{babel}
\usepackage[T1]{fontenc}
\usepackage{mathrsfs}
\usepackage{hyperref}
\usepackage{physics}
\usepackage{mathtools}
\usepackage{multirow}
\usepackage{bm, epsfig}
\usepackage{graphicx}
\usepackage{subeqnarray}
\usepackage{bbold}
\usepackage{upgreek}
\usepackage{tikz}
\usepackage{slashed}
\usepackage{lineno}
\usepackage{longtable}
\usepackage{booktabs}
\usepackage{float}

\hypersetup{colorlinks=true, urlcolor=blue, citecolor=blue, filecolor=blue, linkcolor=blue}

\newcommand{\Li}[2]{\mathrm{Li}_{#1}\left(#2\right)}

\begin{document}

\title{$^{208}$Pb nuclear charge radius revisited: closing the  fine-structure-anomaly gap}

\date{\today}

\author{Zewen~Sun}
\affiliation{Max Planck Institute for Nuclear Physics,  Saupfercheckweg 1, 69117 Heidelberg, Germany}
\author{Konstantin~A.~Beyer}
\affiliation{Max Planck Institute for Nuclear Physics,  Saupfercheckweg 1, 69117 Heidelberg, Germany}
\author{Zoia~A.~Mandrykina}
\affiliation{Max Planck Institute for Nuclear Physics,  Saupfercheckweg 1, 69117 Heidelberg, Germany}
\author{Igor~A.~Valuev}
\affiliation{Max Planck Institute for Nuclear Physics,  Saupfercheckweg 1, 69117 Heidelberg, Germany}
\author{Christoph H. Keitel}
\affiliation{Max Planck Institute for Nuclear Physics,  Saupfercheckweg 1, 69117 Heidelberg, Germany}
\author{Natalia~S.~Oreshkina}
\email[Email: ]{Natalia.Oreshkina@mpi-hd.mpg.de} 
\affiliation{Max Planck Institute for Nuclear Physics,  Saupfercheckweg 1, 69117 Heidelberg, Germany}

\begin{abstract}
A comprehensive reevaluation of the root-mean-square nuclear charge radius is presented for the doubly magic $^{208}$Pb extracted from muonic spectroscopy measurements.
By integrating rigorous theoretical quantum electrodynamics calculations, state-of-the-art numerical methods, and a systematic reanalysis of the uncertainties, we reduced the long-standing muonic fine-structure anomaly and improved the  goodness of fit by a factor of twenty. 
The resulting value of 5.5062(5)~fm for a Fermi distribution is fairly consistent with the previously reported muonic spectroscopy value, and three standard deviations larger than the commonly used compilation data, which indicates  that the current  value and its uncertainty  could be significantly underestimated. 
Attributing the remaining discrepancy to theory errors which can not be rigorously calculated we suggest the rms charge radius with reduced model dependence to be 5.5062(17) fm.
This work sets an improved benchmark for charge radius extraction in heavy nuclei and paves a path for systematic  reevaluations across the nuclear chart. 
\end{abstract}

\maketitle


{\it Introduction. --- }
Nuclear root-mean-square (rms) charge radii serve as fundamental benchmarks bridging various fields of physics. 
They are indispensable input parameters for nuclear-, atomic-, and molecular-physics calculations.  
Reliable rms radii are crucial for precision tests of quantum electrodynamics (QED) and for the determination of fundamental constants. 
Additionally, many searches for physics beyond the Standard Model rely on high-precision rms charge radii.

Accurate calculations of nuclear parameters from first principles are particularly challenging for heavy nuclei. 
Since direct measurements of nuclear charge radii are not possible, indirect extraction with significant theory input is necessary.
Muonic atoms are highly sensitive to these nuclear parameters, such that the most accurate method to measure rms charge radii is based on muonic atom spectroscopy \cite{Wheeler1949,BorieRinker1982}. 
Motivated by the success of light muonic atom spectroscopy~\cite{Parthey2010Precision, Pohl2010Laser, Schuhmann2023Helion} and  new measurements of heavy muonic atoms with microtargets \cite{Antognini2020}, we started the development of a significantly more accurate theory for the  rms radii extraction of heavy nuclei.

While muonic atom spectroscopy provides the most precise determinations of rms charge radii, its application to heavy nuclei has revealed a persistent discrepancy, namely the muonic fine-structure anomaly~\cite{Yamazaki1979,  Phan1985, Bergem1988, Piller1990}, first observed nearly 50 years ago but never resolved.
The theoretically predicted spectra of lead, tin, and zirconium exhibit small but systematic deviations from the experimentally measured values, for any value of the assumed nuclear rms charge radius. 
This discrepancy was attributed to the least understood and most complicated of all contributions: nuclear polarization. 
However, our rigorous microscopic calculations of the nuclear polarization have shown that it is most likely not solely responsible for the anomaly~\cite{Valuev2022}. 
This aligns with the poor fit quality addressed above and hints at the uncertainty of the nuclear radius of double-magic $^{208}$Pb being  significantly underestimated. 
The same situation applies to other muonic atoms, as the methods for rms charge radius extraction have been the same.

\begin{figure}
    \centering
\includegraphics[width=0.9\columnwidth]{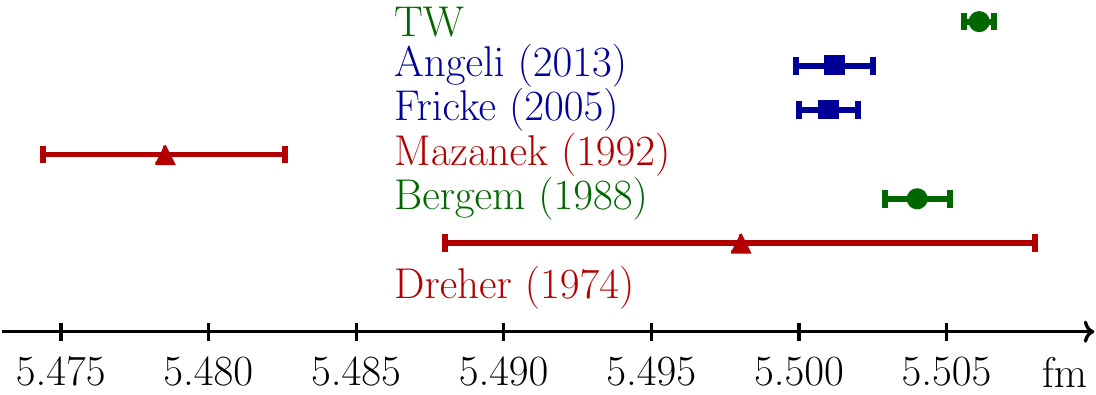}
    \caption{Nuclear rms radius for $^{208}$Pb extracted in this work compared to the literature values. The data based on muonic spectroscopy are marked with green (this work (TW) and Bergem~\cite{Bergem1988}), data based on the electron scattering are in red (Mazanek~\cite{Mazanek1992}\footnote{We actually found this thesis, read and digitalized it.} and Dreher~\cite{DREHER1974}), and the combined values are in blue (Angeli~\cite{Angeli2013Table} and Fricke~\cite{Fricke2004}).     }
\label{fig:rms_comparison}
\end{figure}

In this Letter, we report a theoretical and computational breakthrough towards resolving this anomaly for $^{208}$Pb by combining fully rigorous QED calculations with a modern statistical treatment of uncertainties. We reanalyzed the spectrum of muonic $^{208}$Pb and found a best-fit rms of 5.5062(5) fm for a 2-parameter Fermi (2pF) nuclear charge distribution, which is 3-4$\sigma$ larger than the tabulated value \cite{Angeli2013Table, FRICKE1995177}, see Fig.~\ref{fig:rms_comparison}. 
In contrast to the previously reported and tabulated data, our value is based on fully rigorous {\it ab initio} theory and  improves the goodness of fit by a factor of twenty. 
This significantly reduces  the muonic fine-structure anomaly to the point where its' existence should be questioned.
Therefore we are confident that the here presented rms charge radius of $^{208}$Pb should be adopted as the new standard as extracted from muonic spectroscopy. Our analysis strongly suggests that the currently accepted rms charge radius is too small and must be updated correspondingly.

{\it Theory improvements. --- }
We calculated all individual contributions to the muonic spectrum  which contribute $\gtrsim$10\% of the experimental precision.
The leading contribution is the finite nuclear size~(FNS) effect, obtained by numerically solving the Dirac equation with a nuclear potential derived from a two-parameter Fermi (2pF) nuclear charge model (see Sup.~Mat. for details). 

\begin{table*}
    \begin{tabular}
{p{1.2cm}p{1.7cm}p{1.2cm}p{1.2cm}p{1.2cm}p{1.2cm}p{1.2cm}p{1.5cm}p{1.6cm}p{1.5cm}p{1.3cm}}
    \hline\hline
    State &	2pF binding\, energy & Ue & WK & KS & $\mu$Ue & HadUe & SE & NP & RRecoil & Screening \\
    \hline
    $1s_{1/2}$ & 10525.905 & 67.084 & -0.504 & 0.552 & 0.238 & 0.159 & -3.345(50) & 5.712(603) & -2.555(66) & 5.555  \\ 
    $2s_{1/2}$ & \phantom{1}3581.659 & 19.336 & -0.250 & 0.149 & 0.041 & 0.027 & -0.653(6) & 1.063(156) & -1.317(27) & 5.536 \\
    $2p_{1/2}$ & \phantom{1}4783.784 & 32.290 & -0.356 & 0.251 & 0.044 & 0.030 & -0.440(20) & 1.923(263) & -2.114(4) & 5.548 \\ 
    $2p_{3/2}$ & \phantom{1}4601.766 & 29.761 & -0.344 & 0.229 & 0.033 & 0.023 & -0.745(10) & 1.864(256) & -2.172(2) & 5.547 \\ 
    $3p_{1/2}$ & \phantom{1}2129.324 & 10.759 & -0.167 & 0.081 & 0.014 & 0.009 & -0.180(15) & 0.584(99) & -1.003(3) & 5.512 \\
    $3p_{3/2}$ & \phantom{1}2082.436 & 10.230 & -0.163 & 0.077 & 0.011 & 0.008 & -0.245(10) & 0.619(96) & -1.025(0) & 5.510 \\
    $3d_{3/2}$ & \phantom{1}2163.565 & 10.502 & -0.192 & 0.076 & 0.001 & 0.001 &  \phantom{-}0.020(2) & 0.238(56) & -1.162(1) & 5.525 \\ 
    $3d_{5/2}$ & \phantom{1}2121.477 & \phantom{1}9.848 & -0.185 & 0.071 & 0.001 & 0.000 & -0.070(5) & 0.037(37) & -1.143(0) & 5.523 \\ 
    \hline\hline
    \end{tabular} 
    \caption{ Corrections to muonic $^{208}\mathrm{Pb}$ binding energies (keV) calculated perturbatively with FNS wavefunctions from the 2pF model for the best-fit values $a=0.5283$ fm, $c=6.6409$ fm, $\text{rms radius}=5.5062$ fm. 
    A positive correction refers to an increase in binding energy. }
    \label{tab:corrections}
\end{table*}

To the first order in the fine-structure constant $\alpha$, there are two  QED corrections: vacuum polarization (VP) and self energy (SE). 
In contrast to electronic systems, where these two contributions have similar magnitude, VP dominates in muonic atoms. 
VP can be expressed as a sum of Uehling (Ue) and Wichmann-Kroll (WK) contributions, represented through their respective potentials. 
The dominant Ue corrections are calculated rigorously for a 2pF model using the method~\cite{Klarsfeld1997Analytical}. 
Since numerical precision is the only limiting factor, we ensured that the error is negligible compared to the experimental uncertainty.
For the evaluation of the WK contribution including the FNS effect, two independent methods have been used \cite{Ivanov2024wk,  Zoia_2025}. 
Moreover, we have also included the K\"all\'en-Sabry (KS) contribution, which accounts for the second-order correction beyond the Uehling term, following the method described in Ref.~\cite{Indelicato2013Nonperturbative}. 

In addition to the electron-positron loop, VP can also originate from any other particle-antiparticle (e.g.~muon-antimuon) loop. 
However, these effects are exponentially suppressed by the mass of the virtual particles; thus, going beyond leading order is only necessary for electronic Ue at the current level of accuracy. 
The muon-antimuon loop gives the muonic Ue ($\mu$Ue) correction, which can be calculated {by a straightforward replacement of the virtual particle mass}. 
For the hadronic Ue (HadUe) contribution, we employed the approach~\cite{Breidenbach_PhysRevA.106.042805} derived from hadronic scattering data, and adapted it for muonic atoms~\cite{Moritz_2025}.

The SE correction has been improved compared to Ref.~\cite{Bergem1988} and calculated within a fully rigorous QED framework, following  Ref.~\cite{Oreshkina2022}.
Furthermore, we investigated the model dependence of SE, which resulted in increased uncertainties compared to the previous results.
Similarly, for the relativistic recoil (RRecoil) correction, we adopted the approach of Ref.~\cite{Yerokhin2023recoil} and improved it by incorporating the 2pF model instead of the previously used exponential model. 

Finally, the most cumbersome and challenging correction, the nuclear-polarization (NP) effect, has been calculated in the same manner as {described} in Ref.~\cite{Valuev2022}, now for all states appearing in the spectra.
The uncertainties have been conservatively estimated as the spread of the results for the nine different Skyrme parametrizations that have been used in Ref.~\cite{Valuev2022} and shown to cover a wide range in the parameter space. 
Special care had to be taken in the case of the $3d_{3/2}$ and $3d_{5/2}$ levels due to the resonances between the muonic transitions $3d_{3/2} \rightarrow 2p_{3/2}$, $3d_{5/2} \rightarrow 2p_{3/2}$ and $3d_{5/2} \rightarrow 2p_{1/2}$ and the lowest octupole nuclear excitation~\cite{Shakin_resonances}, see Sup.~Mat. for further information.

In our calculations, we also evaluated the screening effect of the interaction between the muon and the outer surrounding electrons following Ref.~\cite{Michel2017Theoretical}, assuming the lowest electronic shell is filled as $(1s)^2$.
Despite its relatively large absolute impact on individual energy levels, the screening effect remains constant across the low part of the muonic spectrum.
Therefore, the number of remaining electrons and the exact electronic configuration do not affect the spectral lines of heavy muonic systems relevant for the rms radius evaluation.    

In Table \ref{tab:corrections}, we present all corrections, for simplicity  calculated perturbatively {(see Sup. Mat. for details)} on the wavefunctions with FNS from 2pF, to illustrate the hierarchy of the effects. 
For the fitting procedure, all VP effects have been included in the Hamiltonian and, therefore, treated non-perturbatively.

{\it Experimental data. --- }
For our current evaluation, we have used the experimental measurements of $\mu-^{208}$Pb reported in Ref.~\cite{Bergem1988} (see Table~\ref{tab:exp}) and recently confirmed them to be reliable~\cite{Andreas2025}. 
The data are split in three runs, which correspond to the high-, medium-, and low-energy parts of the spectrum. 
This split was arranged to allow for specific tuning of the detector. 
We performed the fit with respect to all transitions, but also analyzed the results from individual runs. 

\begin{table}[h!]
    \centering
    \begin{tabular}{p{2cm} p{3cm} p{2cm}}
    \hline\hline
    Set & Transition &	Energy (keV)  \\
    \hline
    Run I & $2p_{3/2} - 1s_{1/2}$ & 5962.854(90)\\ 
    & $2p_{1/2} - 1s_{1/2}$ & 5778.058(100)\\[1.5mm]
    Run II & $3d_{3/2} - 2p_{1/2}$ & 2642.332(30)\\ 
    & $3d_{5/2} - 2p_{3/2}$ & 2500.590(30)\\ 
    & $3d_{3/2} - 2p_{3/2}$ & 2457.569(70)\\[1.5mm] 
    Run III & $3p_{3/2} - 2s_{1/2}$ & 1507.754(50)\\ 
    & $3p_{1/2} - 2s_{1/2}$ & 1460.558(32)\\ 
    & $2s_{1/2} - 2p_{1/2}$ & 1215.330(30)\\ 
    & $2s_{1/2} - 2p_{3/2}$ & 1030.543(27)\\ 
    \hline\hline
    \end{tabular}
    \caption{Experimentally measured transitions with their error bars~\cite{Bergem1988}.  }
    \label{tab:exp}
\end{table}

Finally, it is important to note that the most recent spectral measurement of muonic lead was performed almost 40 years ago, and there continues to exist tension between theory and the experiment. 
With our rigorous QED calculation we aim to prompt a new experimental campaign to measure the spectrum of muonic lead with a focus on state-of-the-art uncertainty analysis explicitly splitting statistical and systematical errors.

{\it Fitting procedure. --- } 
We calculated the transition energies as a function of 2pF parameters $a$ and $c$, including all corrections described above.
The best-fit parameters are obtained by minimizing $\chi^2$ between the experimental data and the theory prediction. 
The theoretical uncertainties arise from a combination of model dependencies, deviations between different numerical methods, and empirical estimations of unaccounted higher-order contributions,  and are fully taken into account. 
Experimental uncertainties are assumed to be uncorrelated  as shown in Table~\ref{tab:exp}. 
For more detailed information, see Sup.~Mat.

\begin{figure*}
    \centering\includegraphics[width=15cm]{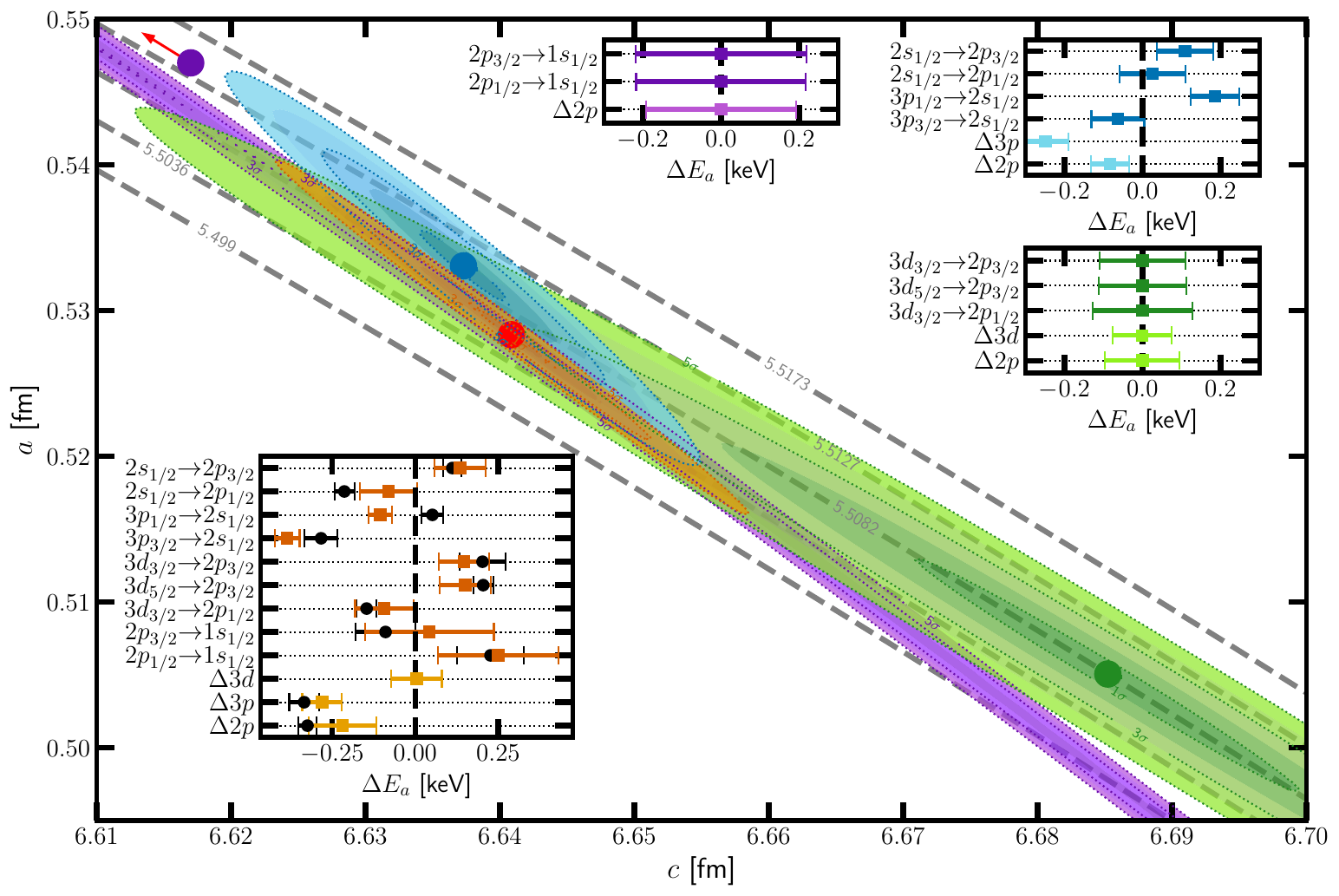}
    \caption{The fit of the spectrum of $\mu-^{208}$Pb to a 2pF distribution. The colored regions show the $5\sigma$ likelihood for the 2pF parameters and the dotted contours highlight the $5$, $3$, and $1\sigma$ steps. Each colour corresponds to a fit of different groups of transitions: purple only fits Run I, green corresponds to Run II, blue to Run III, and orange is the combined fit of all the transitions. The insets show the resulting residual transition energies in the same colours, with measured transitions in darker shade and resulting fine-structure splittings  in lighter colour. The previous fit as taken from Ref.~\cite{Bergem1988} is shown in black wherever available, and the error bars include experimental errors only as the theory uncertainty was not explicitly quoted. The dashed lines correspond to curves of constant radius. The coloured circles indicate the best-fit parameters and the red circle the final values for the overall fit. Note that the best-fit value for Run I lies outside of the plotted parameter range, as indicated by the red arrow.}
    \label{fig:ac_rms}
\end{figure*}

{\it Results and Discussion. --- }
{The best-fit 2pF parameters $c$ and $a$, together with the corresponding rms charge radius,} are shown in Table~\ref{tab:res}.
Fig.~\ref{fig:ac_rms} depicts the likelihood contours for the 2pF  parameters up to the $5\sigma$ level. 
The fit was performed for each of the 3 runs separately and the final value was extracted from a combined fit to all the experimental data. 
The individual fits are represented by different colours: purple, green, and blue correspond to runs I, II, and III, respectively. The full fit is depicted in orange. 
The insets show the residual transition-energy differences between best-fit theory and measurement for each of the fits.
The ellipses corresponding to the 3 individual runs are in mild tension with each other which is ultimately responsible for the poor fit quality of the full fit ($\chi^2 / \mathrm{DoF} = 9.5$). Interestingly, the best-fit values for the individual runs still fall closely to a line of constant rms charge radius.  
Note that the figure does not show the best-fit value for Run I as it is far off the other two, but can be found in Table~\ref{tab:res}.

The likelihood contours for all the fits are well aligned with the lines of constant radius, which explains the relatively small uncertainty on the rms charge radius compared to the individual 2pF parameters. Interestingly, Run I, shown in purple, has a narrow but very extended likelihood surface, which results in a comparatively larger error of the rms charge radius than for the lower-energy runs. This is opposite to the usual notion that the high-energy transitions are most sensitive to the rms charge radius and can be explained by the large theory uncertainty for these low-lying states. We therefore believe that all experimental data should be utilized for the fitting procedure.

To understand the implications of the present analysis it is important to remember the three basic assumptions which underlie the present work:
\begin{itemize}
    \item[(i)] The experimental values including their uncertainties are correct.
    \item[(ii)] We have correctly accounted for every relevant QED effect and  the theory uncertainties.
    \item[(iii)] The nuclear charge distribution is close enough to a 2pF model.
\end{itemize}
We lack the expertise and raw data to comment on the validity of (i). 
We have assessed assumption (ii) to the best of our knowledge. 
It is well known that muonic spectroscopy is very sensitive to nuclear properties, and thus we see inaccuracy of (iii) as the most likely explanation for the remaining discrepancy between individual runs. 

Even though the theory uncertainties have been estimated to the best of our ability, the quality of those estimates can not be checked directly. Assuming that the experimental data is correct and the spectrum is fully described within QED we attribute the remaining discrepancy to unaccounted uncertainties like, e.g. model dependence. 
By normalizing the error such that the $\chi^2/\rm{DoF}$ becomes unity, and ensuring that the spectrum is reproduced within the accuracy of the calculation,  we arrive at a conservative, reduced model-dependent rms charge radius of $5.5062(17)$ fm. 

{\it Comparison with previous data. --- } 
Comparison of our results with the most recent full extraction of the rms charge radius of lead in \cite{Bergem1988} is straightforward. 
The underlying data is the same; however, we have updated QED theory and error analysis. 
We find a best-fit rms charge radius which is in mild tension with their previously reported muonic spectroscopy value, see Table \ref{tab:res}. 

The results of Bergem~\cite{Bergem1988} were combined with electron scattering data and optical isotope-shift measurements in the compilations of Fricke \cite{FRICKE1995177} and Angeli \cite{Angeli2013Table}. 
This combination of data sets is performed to empirically correct for the assumption of a nuclear 2pF distribution. 
Taken at face value, their final number is in strong tension with our result. 
We discuss below that the systematic error introduced through the exact shape of the nuclear charge density can not be quantified exactly; however we still believe that the tabulated rms charge radius of lead requires updating.

{\it {Model dependence}. ---} 
Based on the recent indication that ${}^{208}$Pb is slightly deformed \cite{Henderson2025Deformation}, we investigated the effect on the rms radius value.
We calculated the corresponding deformation parameter~$\beta$, following the methods described in Sup.~Mat., and performed the fit for the deformed Fermi model. 
The resulting rms radii of the two models are the same, up to all significant figures reported in Table~\ref{tab:res}. 
The difference in $\chi^2 / \mathrm{DoF}$ between the two models is also negligible. 

As a basic test of model dependence we also considered a 3pF model with additional free parameter $w$ (see Sup. Mat.). 
The resulting fit is consistent with the 2pF model and the reduced $\chi^2/\mathrm{DoF}$ does not suggest preference for this modification. 
We would like to stress that a full analysis of the model dependence would require a parametrization of all possible density functions and is therefore not feasible until new, {\it ab initio} nuclear densities are available.

In the absence of `better' test-distributions, rooted in physical principles, the exact level of accuracy at which the deviation of the nuclear charge distribution from 2pF affects the rms charge radius can only be estimated. 
A conservative limit would be to compare the 2pF specific value from muonic spectroscopy with values extracted from electron-scattering experiments. 
The latter are, in principle, sensitive to the shape of the charge distribution; however, the subtraction of QED effects faces challenges. 
As we lack the expertise to comment on the procedure, we simply highlight that the rms charge radii tend to deviate on the level of $\sim$0.001 fm.

\begin{table}[h!]
    \centering
    \begin{tabular}{p{0.9cm}p{1.3cm}p{1.3cm}p{1.3cm}p{1.5cm}p{1.5cm}}
    \hline\hline
     & Run I & Run II & Run III & Total & Ref.~\cite{Bergem1988} \\\hline
    $c$ & 6.55(17) & 6.685(66) & 6.639(7) & 6.6409(25) & 6.6447(5) \\ 
    $a$ & 0.59(11) & 0.505(34) & 0.531(5) & 0.5283(18) &  0.5249(3) \\ 
        rms  & \multirow{ 2}{*}{5.525(46)} & \multirow{ 2}{*}{5.508(5)} & \multirow{ 2}{*}{5.509(2)} & \multirow{ 2}{*}{5.5062(5)} & \multirow{ 2}{*}{5.5040(11)}\\[-4pt]
        radius &&&&&\\
    $^{\chi^2} / _{\mathrm{DoF}}$ & \, --- & 0.0001 & 2.9 &  9.5&  187 \\ 
    \hline\hline
    \end{tabular}
    \caption{Nuclear parameters of $^{208} \mathrm{Pb}$ determined from the fits to muonic transition energies, in fm. }
    \label{tab:res}
\end{table}

{\it Conclusions. ---} 
We performed fully rigorous, state-of-the-art QED calculations for the spectrum of muonic lead to extract the nuclear rms charge radius from a fit to experimental measurements.
Using a two-parameter Fermi distribution, we determine the nuclear rms radius value of 5.5062(5) fm, achieving a twentyfold improvement in fit quality over previous analyses.
The remaining overall high $\chi^2/{\rm DoF} = 9.5$ we traced back to an inconsistency between the fits of groupings of the transitions by energy.

Our analysis reduces the previously reported  fine-structure anomaly in $\Delta 2p$ from 6$\sigma$ to 2$\sigma$, suggesting that the apparent anomaly arose primarily from underestimated uncertainties in earlier analyses.
Contrary to conventional understanding, our analysis demonstrates that transitions involving $2s$, $2p$ and $3p$ states impose at least equally stringent constraints on radius extraction as traditionally emphasized $2p - 1s$ transitions. This observation may influence future experimental strategies.

To reduce residual model dependence, we conservatively normalize uncertainties to achieve $\chi^2/\mathrm{DoF} = 1$, yielding a reduced model-dependent  estimate of 5.5062(17)~fm. This value is consistent with the previous work in Ref.~\cite{Bergem1988}, which analyzed the same muonic spectral data, but our work incorporates complete QED corrections for the first time.
Our result is significantly larger than the currently accepted standard based on combined muonic and electron-scattering data.

We are convinced that the current standard rms charge radius of $^{208}$Pb may be underestimated and requires reconsideration in light of this work. More broadly, our results strongly indicate the necessity of a systematic, theory-driven reevaluation of the rms charge radii across the periodic table along the avenue demonstrated here using muonic spectroscopy and modern QED theory.

{\it Acknowledgements. --- }
The Authors thank A.~Knecht, F. Hei{\ss}e, B.~Ohayon, V.~A.~Yerokhin, and V.~A.~Zaytsev for discussions.
This work is part of and funded by the Deutsche Forschungsgemeinschaft (DFG, German Research Foundation) under the Collaborative Research
Centre, Project-ID No. 273811115, SFB 1225 ISOQUANT.
This article comprises parts of the PhD thesis work of Z.~M. and Z.~S. to be submitted to Heidelberg University.

\bibliography{refs}

\begin{widetext}

\section{Supplemental Material}
\end{widetext}

Relativistic units ($m_e = \hbar = c = 1$) and the Heaviside charge units ($e^2 = 4\pi\alpha$) are used below, unless explicitly given.

\section{Physical constants} 
The physical constants used in this work are listed in Table~\ref{tab:constants}. 
All values are retrieved from 2022 CODATA~\cite{codata2022}, except the $^{208}$Pb mass, which is obtained from Ref.~\cite{Audi2003Ame}, and is used in the recoil correction calculation. 

\begin{table}[h]
    \centering
    \begin{tabular}{ll}
    \hline\hline
    fine-structure constant$^a$ & $\alpha = 1/137.035999177 $ \\
    electron mass energy$^a$ & $m_e c^2 = 0.51099895069 $ MeV \\
    muon mass energy$^a$ & $ m_\mu  c^2 = 105.6583755 $ MeV \\
    reduced Planck constant$^a$ & $\hbar c = 197.3269804$ MeV$\cdot$fm \\
    $^{208}$Pb  mass$^b$ & $m_\mathrm{Pb} = 207.9766521$ u \\
    atomic mass unit$^a$ & $(1\mathrm{u})c^2$ = 931.49410372 MeV \\
    \hline\hline
    {$a$ Adopted from CODATA \cite{codata2022}}\\
    {$b$ Adopted from Ref.~\cite{Audi2003Ame} }
    \end{tabular}
    \caption{Physical constants used in this work.}
    \label{tab:constants}
\end{table}

\section{Fermi charge distribution}\label{SupMat:fermi}

The two-parameter Fermi (2pF) distribution is a common theoretical model for nuclear charge distribution
\begin{equation}
    \rho_\mathrm{2pF}(r) = \frac{ \rho_0 }{1+ \exp[(r-c)/a]}, 
\end{equation}
where $a$ and $c$ are two parameters for 2pF, and $\rho_0$ is the normalization factor 
\begin{equation}
    \rho_0=\frac{1}{-8\pi a^3\Li{3}{-\exp\left(c/a\right)}},
\end{equation}
with $\Li{n}{x}$ being a polylogarithm of order $n$. 
The corresponding nuclear potential can be calculated as
\begin{equation}
    V_\mathrm{nucl}(r) = -\alpha Z \int{\rm d}^3\mathbf{r}'\frac{\rho(r')}{\max(r,r')}.
\end{equation}

The corresponding rms radius can be expressed as 
\begin{equation}
\label{eq:rms_full}
     r_\mathrm{rms}^2 = 12 a^2\frac{\Li{5}{-\mathrm{exp}\left(c/a\right)}}{\Li{3}{-\mathrm{exp}\left(c/a\right)}},
\end{equation}
and the uncertainty of $r_\mathrm{rms}$ is defined as
\begin{equation}
     \sigma_{ r } = \sqrt{ \abs{ \frac{ \partial r}{ \partial c} }^2 \sigma_c^2 + \abs{ \frac{ \partial r}{ \partial a} }^2 \sigma_a^2 + 2 \frac{ \partial r}{ \partial c} \frac{ \partial r}{ \partial a} \sigma_{ca} }, 
\end{equation}
where $\sigma_c$ and $\sigma_a$ are uncertainties of $c$ and $a$, respectively, and $\sigma_{ca}$ is the covariance of $c$ and $a$. 

A recent measurement \cite{Henderson2025Deformation} suggests that  the $^{208}$Pb nuclear shape is slightly deformed, which is depicted by the deformed Fermi (defF) model
\begin{equation}
    \rho_{\mathrm{defF}} (r, \theta) = \frac{ \rho_0 }{1+ \exp \left( \left\{ r-c [ 1+\beta Y_2^0(\theta) ] \right\} / a \right) }, 
\end{equation}
where $\beta$ is the deformation parameter, and $Y_l^m(\theta)$ are spherical harmonics. 
The measured nuclear electric quadrupole transition rate \cite{Henderson2025Deformation} is
$$ B(E2;2^{+} \rightarrow 0^{+}) = 6.5 \ \mathrm{W.u.},$$ 
where $\mathrm{W.u.}$ stands for Weisskopf units~\cite{Wu2025}. 
This rate can be used to determine the deformation parameter $\beta$ using the procedure described in Ref.~\cite{Sun2024Nuclear}, or via an approximate formula of
\begin{equation}
\label{eq:beta_approx}
    \beta \approx \frac{4 \pi }{ 5 Z \, r_\mathrm{rms}^2}  \sqrt{ \frac{B(E2\uparrow)}{e^2} }. 
\end{equation}
where 
\begin{equation}
    B(E2\uparrow) = \frac{2J + 1}{2J_0 + 1} \, B(E2\downarrow), 
\end{equation}
with $J_0$ and $J$ being spins of the lower and upper levels, respectively.

For $^{208}$Pb, according to the measurement~\cite{Henderson2025Deformation}, Eq.~\eqref{eq:beta_approx} gives $\beta = 0.049$, whereas the exact approach~\cite{Sun2024Nuclear} provides $\beta = 0.052$. 
We used $\beta=0.05$ to perform the fit for the defF model, and found no difference in the resulting rms radii, up to all significant figures, as shown in Table~\ref{tab:res_model}. 

Four-parameter Fermi (4pF) distributions were introduced in order to match the nuclear charge distribution obtained from the electron-scattering experiments~\cite{Heisenberg1969Elastic}
\begin{equation}
    \rho_\mathrm{4pF}(r) = \frac{ \rho_0 \left[1 + w \left(r^2/c^2\right) \right] }{1 + \exp[(r^n - c^n)/a^n]}. 
\end{equation}
Is is also known as 3pF model with parameters $a$, $c$, $w$, and $n=1$. 
We present the best-fit parameters of 3pF and 4pF models in Table~\ref{tab:res_model}. 
Promoting 2pF to more sophisticated models can slightly improve the overall agreement between theoretical and experimental transitions, with slightly different rms radius values. 
However, the goodness of fit, quantified by $\chi^2 / \mathrm{DoF}$, is not improved, since the extra parameters increased the number of degrees of freedom~(DoF). 
It should also be noted that since the fitting procedure can be trapped at local minima, it is hard to guarantee the presented parameters to be the global minimum points of $\chi^2$ in the parameter space of 3pF and 4pF.

\begin{table}[h!]
    \centering
    \begin{tabular}{p{1.8cm}p{1.5cm}p{1.5cm}p{1.5cm}p{1.5cm}}
    \hline\hline
    Parameter & 2pF & defF & 3pF & 4pF \\\hline
    $c$ & 6.6409 & 6.6405 & 6.4356 & 6.2960 \\ 
    $a$ & 0.5283 & 0.5261 & 0.5557 & 2.8938 \\ 
    $\beta$ & -- & 0.05 & -- & -- \\
    $w$ & -- & -- & 0.2314 & 0.3386 \\
    $n$ & -- & -- & 1 & 1.9967 \\
    rms & 5.5062 & 5.5062  & 5.5075 & 5.5056 \\
    $\chi^2$ & 66.8 & 66.8 & 65.2 & 68.7 \\ 
    $^{\chi^2} / _{\mathrm{DoF}}$ & 9.5 & 9.5 & 10.9 & 13.7 \\ 
    \hline\hline
    \end{tabular}
    \caption{Nuclear parameters of $^{208} \mathrm{Pb}$ determined from different nuclear distribution models. For the definition of $\chi^2$, please see Eq.~\eqref{eq:chi} and the text below. }
    \label{tab:res_model}
\end{table}

\section{QED contributions}
\begin{figure}[h]
    \centering
    \includegraphics[width=0.99\linewidth]{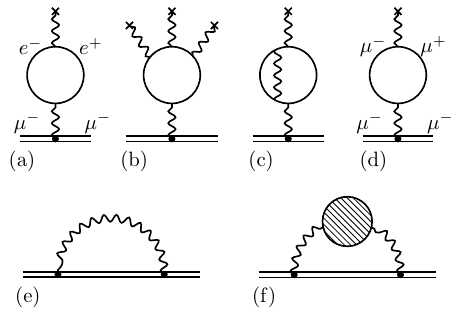} 
    \caption{Feynman diagrams for individual contributions: (a)~Uehling, (b)~Wichmann-Kroll, (c)~K\"all\'en-Sabry, (d)~muonic Uehling, (e)~self energy and (f)~nuclear polarization.}
    \label{fig:QED_contributions}
\end{figure}

Feynman diagrams for some calculated effects are listed in Fig.~\ref{fig:QED_contributions}. 

Every contribution originated from vacuum-polarization (VP) correction to the energy levels can be represented in the form of a corresponding potential. 
These potentials can be summed and added to the initial potential in the Dirac equation. 
However, the resulting energy correction from the solution of the Dirac equation does not inherit this linearity and is represented by a rather complex function of individual contributions. 
Therefore, the energy corrections in Table~I of the main text are calculated to the first order of perturbation theory, for specific potentials, e.g.:
\begin{equation}
    \Delta E_{{\rm Ue},A} = \bra A V_{\rm Ue} \ket A,
\end{equation}
which is calculated for the state {$A$} {using} the wavefunctions obtained by numerical solutions of the Dirac equation with a 2pF nuclear charge distribution.

In the fitting procedure, the energies have been calculated by including the potentials of Uehling~(Ue), Wichmann-Kroll~(WK), K\"all\'en-Sabry~(KS), muonic Uehling~($\mu$Ue), and hadronic Uehling~(HadUe) into the Dirac equation on the level of Hamiltonian. 
Nevertheless, self energy~(SE), nuclear polarization~(NP), and relativistic recoil~(RRecoil) can not be represented via potentials. 
Therefore, the corresponding contributions were added to the total binding energy perturbatively. 

\section{Resonances in the NP effect}
The term ``muon-nuclear resonance'' refers to effects coming from quasi-degenerate product states in the combined muon-nucleus basis.
Such effects can play a crucial role for fine-structure splittings even when the distances between the corresponding muon-nucleus product states are of the order of 100 keV and the perturbation theory is still perfectly applicable.
This is the case for the $\Delta 3d$ splitting and the lowest octupole nuclear excitation $|3^{-}\rangle$ in muonic Pb, where the $3d_{3/2}$ muonic level comes close in energy to the $|2p_{3/2}\rangle \otimes |3^{-}\rangle$ product state, while the $3d_{5/2}$ muonic level turns out to lie closely between the $|2p_{1/2}\rangle \otimes |3^{-}\rangle$ and the $|2p_{3/2}\rangle \otimes |3^{-}\rangle$ product states.
Even though the resulting contributions to the NP corrections amount to only about 10--100 eV and thus are of minor importance for the muonic $3d \rightarrow 2p$ transition energies themselves, they are essential for an accurate description of the more subtle $\Delta 3d$ fine-structure splitting.

In order to take a proper account of these resonances, the following modifications of the NP calculations in the case of the $3d_{3/2}$ and the $3d_{5/2}$ muonic levels were made:
\begin{itemize}\setlength\itemsep{-5mm}
    \item[(i)] The Uehling potential was included in the muonic part of the calculations for a more accurate placement of the unperturbed muonic energy levels. \\
    \item[(ii)] The energy of the lowest $3^{-}$ nuclear excitation was fixed to the experimental value $\omega_{3^{-}} = 2.614522 $~MeV~\cite{Nucl_data_208}, and \\
    \item[(iii)] the corresponding nuclear transition charge-current densities were renormalized to the experimental reduced transition probability $B(E3) = 0.611 \, e^2b^3$~\cite{Nucl_data_208}. 
\end{itemize}

\begin{table*}[!htbp]\label{tab:np_energies}
\caption{Model dependence of NP for individual energy levels (eV).}
\begin{ruledtabular}
\begin{tabular}{ccccccccccc}
State & KDE0 & SKX & SLy5 & BSk14 & SAMi & NRAPR & SkP & SkM* & SGII & avg.(range) \\
\hline\\[-10pt]
$1s_{1/2}$ & -5463 & -5432 & -5557 & -5588 & -5727 & -5889 & -5815 & -5905 & -6035 & -5712(603) \\ 
$2s_{1/2}$ & -1021 & -970 & -1048 & -1030 & -1045 & -1095 & -1116 & -1112 & -1126 & -1063(156) \\[5pt]
$2p_{1/2}$ & -1781 & -1850 & -1834 & -1900 & -1937 & -1997 & -1955 & -2005 & -2044 & -1923(263) \\
$2p_{3/2}$ & -1725 & -1798 & -1776 & -1852 & -1877 & -1936 & -1886 & -1942 & -1981 & -1864(256) \\
$3p_{1/2}$ & -529 & -576 & -556 & -566 & -616 & -540 & -628 & -614 & -627 & -584(99) \\
$3p_{3/2}$ & -559 & -612 & -589 & -602 & -648 & -576 & -672 & -645 & -664 & -619(96) \\[5pt]
$3d_{3/2}$ & -221 & -268 & -219 & -252 & -225 & -212 & -268 & -248 & -230 & -238(56) \\
$3d_{5/2}$ & -23 & -29 & -33 & -21 & -47 & -58 & -40 & -35 & -49 & -37(37) \\[5pt]
\end{tabular}
\end{ruledtabular}
\end{table*}

\begin{table*}[!htbp]
\caption{Model dependence of NP for transition energies (eV).}
\begin{ruledtabular}
\begin{tabular}{ccccccccccc}
Transition & KDE0 & SKX & SLy5 & BSk14 & SAMi & NRAPR & SkP & SkM* & SGII & avg.(range) \\
\hline\\[-10pt]
$2p_{3/2} \rightarrow 1s_{1/2}$ & 3738 & 3634 & 3781 & 3736 & 3850 & 3953 & 3929 & 3963 & 4054 & 3849(420) \\ 
$2p_{1/2} \rightarrow 1s_{1/2}$ & 3682 & 3582 & 3723 & 3688 & 3790 & 3892 & 3860 & 3900 & 3991 & 3790(409) \\[5pt]
$3d_{3/2} \rightarrow 2p_{1/2}$ & 1560 & 1582 & 1615 & 1648 & 1712 & 1785 & 1687 & 1757 & 1814 & 1684(254) \\
$3d_{5/2} \rightarrow 2p_{3/2}$ & 1702 & 1769 & 1743 & 1831 & 1830 & 1878 & 1846 & 1907 & 1932 & 1826(230) \\
$3d_{3/2} \rightarrow 2p_{3/2}$ & 1504 & 1530 & 1557 & 1600 & 1652 & 1724 & 1618 & 1694 & 1751 & 1626(247) \\[5pt]
$3p_{3/2} \rightarrow 2s_{1/2}$ & 462 & 358 & 459 & 428 & 397 & 519 & 444 & 467 & 462 & 444(161) \\
$3p_{1/2} \rightarrow 2s_{1/2}$ & 492 & 394 & 492 & 464 & 429 & 555 & 488 & 498 & 499 & 512(161) \\[5pt]
$2s_{1/2} \rightarrow 2p_{1/2}$ & 760 & 880 & 786 & 870 & 892 & 902 & 839 & 893 & 918 & 860(158) \\
$2s_{1/2} \rightarrow 2p_{3/2}$ & 704 & 828 & 728 & 822 & 832 & 841 & 770 & 830 & 855 & 801(151) \\
\end{tabular}
\end{ruledtabular}\label{tab:np_transitions}
\end{table*}

\section{Fitting procedure}
The rms radius is calculated from Fermi $a$ and $c$ parameters, which are determined from the fit of experimental and theoretical data by minimizing $\chi^2$, which is defined as 
\begin{align}\label{eq:chi}
    \chi^2 &= \Delta^T \cdot \sigma^{-1} \cdot \Delta \\
    \Delta_i &= E_{\mathrm{exp},i} - E_{\mathrm{theo},i}, 
\end{align}
where $E_{\mathrm{exp},i}$ and $E_{\mathrm{theo},i}$ are experimental and theoretical transition energies of the $i$-th transition, $\Delta_i$ defines the components of vector $\Delta$, and $\sigma^{-1}$ is the inverse total covariance matrix. 
In our case, we have 9 transitions (see Table II in the main text), so $\Delta$ has 9 components, and $\sigma$ is a $9 \times 9$ matrix.
The matrix element of $\sigma$ is defined as
\begin{equation}
    \sigma_{ij} = \mathrm{cov} \left( E_{\mathrm{exp},i}, E_{\mathrm{exp},j} \right) + \mathrm{cov}\left( E_{\mathrm{theo},i}, E_{\mathrm{theo},j} \right), 
\end{equation}
where $\mathrm{cov}\left( {E_{i}, E_{j}} \right)$ is the covariance of the $i$-th and $j$-th transition energies. 

We treat the experimental uncertainties $\sigma_{E_{\mathrm{exp},i} }$ reported in~\cite{Bergem1988} as independent and uncorrelated,  such that
\begin{equation}
    \mathrm{cov} \left( E_{\mathrm{exp},i}, E_{\mathrm{exp},j} \right) = \delta_{ij} \sigma_{E_{\mathrm{exp},i} }^2, 
\end{equation}
where $\delta_{ij}$ is the Kronecker delta.
The situation is different for theoretical data, where each state has independent binding-energy uncertainty, whereas calculated transition energies have correlated uncertainties. 
Then, the covariance between transitions $A \rightarrow B$ and $C \rightarrow D$ is
\begin{align}
    \mathrm{cov}( A \rightarrow B, C &\rightarrow D ) = \mathrm{cov}(A,C) - \mathrm{cov}(A,D) \notag \\
    &- \mathrm{cov}(B,C) + \mathrm{cov}(B,D) \notag \\ 
    =& ( \delta_{AC} - \delta_{AD} ) \sigma_A^2 + (\delta_{BD} - \delta_{BC}) \sigma_B^2,
\end{align}
with the corresponding theory uncertainties $\sigma$. 
This method assumes uncertainties of different states to be independent. 
In our case, they are highly correlated because of NP; therefore, we further improved the procedure to minimize the uncertainty, see the next section.

The goodness of fit is measured by $\chi^2 / \mathrm{DoF}$, and the number of degrees of freedom (DoF) equals the number of observations minus the number of free parameters in the fit. 
The smaller $\chi^2 / \mathrm{DoF}$, the better fit quality is. 
In this work, we use, e.g., 9 transitions to fit 2 parameters, $c$ and $a$, and thus we have $\mathrm{DoF} =  7$.

\begin{figure*}
    \centering
    \includegraphics[width=0.9\linewidth]{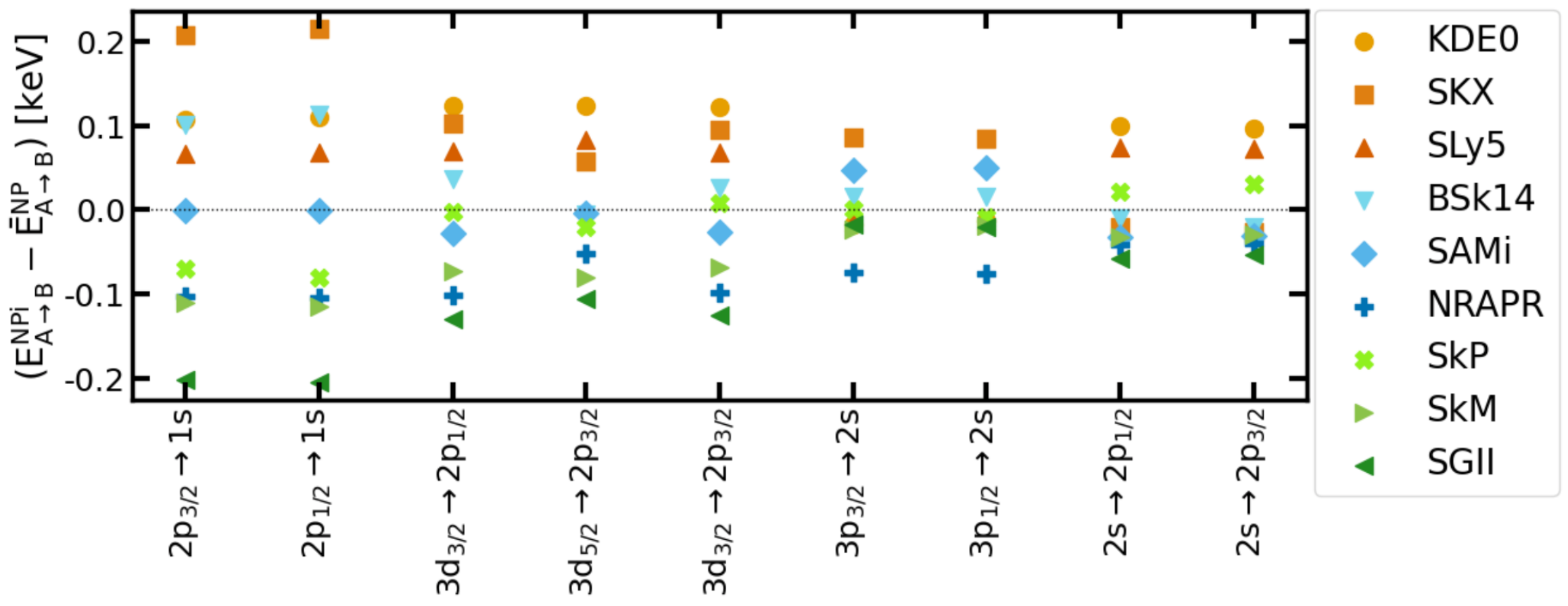}
    \caption{{This plot shows the various NP models listed in Table \ref{tab:np_transitions}. Each model predicts a set of NP corrections to the various transitions. All those are plotted against the mean value here.}}
    \label{fig:NP_models}
\end{figure*}

If the theory prediction fits the experiment, any remaining residual between the two should be consistent with random fluctuations stemming from both the experimental measurement and the theory prediction. Formally this amounts to the hypothesis that the residuals are Gaussian random variables centered around $0$ with covariance $\Sigma$. Under this assumption, the present experiment can be understood as measuring a set of residuals $r_i = E_i^\mathrm{exp} - E_i^\mathrm{Theo}$ from a multivariate Gaussian distribution $\mathrm{N}(\mathbf{r},0,\Sigma)$. The probability of seeing residuals at least as large as we do in the present case is then given by
\begin{equation}
	p = \mathrm{P}(\lvert\mathbf{x}\rvert \geq \lvert\mathbf{r}\rvert) = \int_{\lvert\mathbf{x}\rvert \geq \lvert\mathbf{r}\rvert} \mathrm{N}(\mathbf{x},0,\Sigma).
\end{equation}
For the present fit the probability amounts to $p=6.35\times 10^{-8}$ which should be seen as rejecting the hypothesis that the residuals are Gaussian random noise. It is customary to re-phrase the result as an equivalent significance by comparing it to a measurement of a single Gaussian random variable $x$ and obtaining a result which is equally (un)likely. Under the assumption of Gaussian noise, the present measurement is equivalent to the measurement of a single Gaussian variable and obtaining a result $\gtrsim 5\sigma$ away from the mean value.

This firmly rejects the original assumption and implies that the poor fit is not explained by fluctuations. We must be careful, however, when drawing further conclusions because any number of the underlying assumptions of the fit may be compromised. The most likely of those concerns is the assumption of a 2pF, which introduces a systematic error which is not accounted for in this analysis and not simply quantifiable. 

\section{Uncertainty analysis} \label{uncertainty}

Having a realistic and reliable estimation of uncertainties is important, since  their values have a non-negligible impact on the final result.
NP and SE are the two corrections with the largest uncertainties. 
They can not be included on the level of the potential, but instead are added as separate corrections to the energy levels of the muonic states or transitions.

To estimate the level of uncertainty in NP we follow Ref. \cite{Valuev2022} and take the spread of a set of nuclear Skyrme-type models ($i=1,2,...,N$), listed in Table~\ref{tab:np_transitions}, as a proxy for the theoretical uncertainty of this correction. Hence, we take the average value
\begin{equation}
\label{Eq:NP_average}
    \bar{E}_{A\rightarrow B}^\mathrm{NP} = \frac{1}{N}\sum_{i=1}^N E_{A\rightarrow B}^{\mathrm{NP} i},
\end{equation}
as the NP correction with uncertainty given by
\begin{equation}
\label{Eq:NP_err}
    \left(\sigma_{A\rightarrow B}^\mathrm{NP}\right)^2 = \frac{1}{N-1}\sum_{i=1}^N \left(E_{A\rightarrow B}^{\mathrm{NP} i} - \bar{E}_{A\rightarrow B}^\mathrm{NP}\right)^2.
\end{equation}
The denominator $N-1$ instead of $N$ accounts for the fact that the list of nuclear models included in this work is only a subset of the entire space of models. Because of this, the average value \eqref{Eq:NP_average} of the sample is only an approximation to the average value of the full population of models. 

As can be seen in Fig.~\ref{fig:NP_models}, significant correlation is present between the various transitions and models under investigation. 
Ignoring this correlation could lead to artificially inflated error bars by, e.g., picking alternating upper and lower extreme points for the various transitions, a scenario which is not present in any of the Skyrme force models used. 
By definition, the sample covariance matrix for our list of models is
\begin{align}
\label{Eq:NP_cov}
    \mathrm{cov}&\left(E^\mathrm{NP}_{A\rightarrow B},E^\mathrm{NP}_{ C\rightarrow D}\right) \nonumber\\
    &= \frac{1}{N-1}\sum_{i=1}^N \left(E_{A\rightarrow B}^{\mathrm{NP} i} - \bar{E}_{A\rightarrow B}^\mathrm{NP}\right) \left(E_{C\rightarrow D}^{\mathrm{NP} i} - \bar{E}_{C\rightarrow D}^\mathrm{NP}\right),
\end{align}
where once again, the denominator takes into account the difference between the population mean value and the general mean value of all possible models. 
The diagonal terms reduce to the expression in Eq.~\eqref{Eq:NP_err}. 
Defining the spread in this way makes the implicit assumption that our selection of models is a good representation of all possible models. 
While we have made an effort to include a modern and diverse subset of all Skyrme models we are aware of, we can not be certain that this assumption holds for a general description of the nucleus. 
Therefore, we present the results of the procedure detailed above for a rescaled sample covariance matrix \eqref{Eq:NP_cov} by a factor of~$2$.

For the SE correction, we have estimated the level of uncertainty in the theoretical calculation as a combination of the spread between different models, integration techniques and convergence issues.
Because of the nature of these uncertainties and the observation that they are generally small compared to NP uncertainties, we have limited ourselves to an estimation of the physical correlation arising from the fact that multiple transitions test the same underlying states. Thus, we are restricted to estimating the SE covariance matrix as
\begin{align}
    \mathrm{cov}\left(E^\mathrm{SE}_{A\rightarrow B},E^\mathrm{SE}_{ C\rightarrow D}\right)
    &=\mathrm{cov}\left(E^\mathrm{SE}_A,E^\mathrm{SE}_C\right)+\mathrm{cov}\left(E^\mathrm{SE}_B,E^\mathrm{SE}_D\right)
     \nonumber\\ - \mathrm{cov}&\left(E^\mathrm{SE}_A,E^\mathrm{SE}_D\right)-\mathrm{cov}\left(E^\mathrm{SE}_B,E^\mathrm{SE}_C\right),
\end{align}
where the covariances of the states are diagonal
\begin{equation}
    \mathrm{cov}\left(E^\mathrm{SE}_A, E^\mathrm{SE}_B\right) = (\sigma^\mathrm{SE}_A)^2 \delta_{AB}.
\end{equation}

Finally, relativistic nuclear recoil (RRecoil) also has a sizeable uncertainty originating from nuclear charge density model dependence. The RRecoil correction can be divided into two parts, the lowest-order (LO) one and the remaining, higher-order (HO) part. In \cite{Yerokhin2023recoil} both are evaluated for an exponential nuclear charge distribution, an assumption which proved good enough for electronic systems, but is not suitable for muonic atoms. Fortunately, the LO calculation does not depend on analytic properties of the nuclear charge distribution and therefore can simply be recalculated for a 2pF model. The HO part, unfortunately, can not be  handled the same way. To estimate the uncertainty we take the difference between the two models for the LO contribution and scale it to the absolute size of the HO correction
\begin{equation}
	\delta E_\mathrm{RRec} = \frac{\lvert E_\mathrm{RRec,LO}^\mathrm{exp} - E_\mathrm{RRec,LO}^\mathrm{2pF}\rvert}{E_\mathrm{RRec,LO}^\mathrm{exp}} E_\mathrm{RRec,HO}^\mathrm{exp}.
\end{equation}
The RRecoil uncertainty is then included in analogy to the SE uncertainty.

The final covariance matrix $\boldsymbol{\Sigma}$ of the residuals  $\Delta E = E^\mathrm{QED} - E^\mathrm{exp}$ is calculated as the sum of the individual covariance matrices:
\begin{align}
    \Sigma_{A\rightarrow B,C\rightarrow D} &= \mathrm{cov}\left(\Delta E_{A\rightarrow B},\Delta E_{ C\rightarrow D}\right) \nonumber\\
    &= \mathrm{cov}\left(E_{A\rightarrow B}^\mathrm{NP}, E_{ C\rightarrow D}^\mathrm{NP}\right) + \mathrm{cov}\left(E_{A\rightarrow B}^\mathrm{SE}, E_{ C\rightarrow D}^\mathrm{SE}\right) \nonumber\\
    &\quad + \mathrm{cov}\left( E_{A\rightarrow B}^\mathrm{exp}, E_{ C\rightarrow D}^\mathrm{exp}\right).
\end{align}
The experimental covariance matrix is diagonal, as discussed above.

After redefining the covariance matrix with the inclusion of NP correlations, we can again perform the fit.
The fit results in best-fit values for the 2pF parameters $c=6.6409(25)$ fm and $a=0.5283(18)$ fm. 
These correspond to a rms charge radius of $r_\mathrm{rms} = 5.5062(5)$ fm. The total square error per degree of freedom at the best-fit value is $\chi^2/7 = 9.7$.

\section{Historical values}

\begin{table}[h]
    \centering
    \begin{tabular}{p{3.0cm}p{1.5cm}p{3.5cm}}
        \hline\hline
        Source & Value & Comment \\
        \hline
        Belicard~1967 \cite{Bellicard1967Elastic} & 5.38(3) 5.39(3) 5.42(5) 5.54  & $e$ scattering, obtained from different electron incident energies and 3pF nuclear model \\[2mm]
        Anderson 1969  \cite{Anderson1969Precise} & 5.4978(30) & muonic atoms \\[2mm] 
        Heisenberg 1969 \cite{Heisenberg1969Elastic} & 5.535  5.539 5.546 & $e$ scattering, 3pF and 4pF models \\ & 5.501 5.502 & $e$ scattering + muonic data\\[2mm]
        Dreher 1974 \cite{DREHER1974}  & 5.498(15)  5.520(23) 5.498(10) 5.514(28) & $e$ scattering, different energies and fitting models \\[2mm]
        Kessler 1975 \cite{Kessler1975Muonic} & 5.5097(6) & muonic atoms,  $\chi^2 / N = 0.11$ (see also the text) \\[2mm]     Euteneuer 1976 \cite{Euteneuer1976Charge} & 5.500(24) & $e$ scattering 3pF, \phantom{123} $\chi^2_{\mathrm{min}} \approx 24$\\
        & 5.494(24) & "model independent" \\[2mm]
        Euteneuer 1978 \cite{Euteneuer1978Charge} & 5.4927(75) & $e$ scattering $\overline{\chi^2} \approx 20$\\
        & 5.5032(16) & $e$ scattering + muonic data, $\overline{\chi^2} \approx 22$ \\[2mm]
        Bergem 1988 \cite{Bergem1988} & 5.5040(11) & muonic atoms, \phantom{1234567} $\chi^2/{\mathrm{DoF}} \approx 200$ \\
        \hline\hline
    \end{tabular}
    \caption{A summary of rms radius values from different papers. }
    \label{tab:historical}
\end{table}

In Table~\ref{tab:historical} we summarize previously reported data for the rms radius of $^{208}$Pb obtained from different approaches: electron scattering experiments \cite{Bellicard1967Elastic, Heisenberg1969Elastic, Sick1973Charge, DREHER1974, Euteneuer1976Charge, Euteneuer1978Charge, Mazanek1992}, muonic spectroscopy \cite{Anderson1969Precise, Kessler1975Muonic, Bergem1988}, or combined analysis~\cite{Heisenberg1969Elastic, Euteneuer1978Charge, Fricke2004, Angeli2013Table}. 
We would like to stress that those numbers should probably be taken with a grain of salt, since some of the old data has no error bars at all, and some other values can be underestimated, as they clearly disagree with others obtained by the same methods.
Some of the papers seem to use $\chi^2$ in the sense of $\chi^2/\mathrm{DoF}$ without a clear distinction, and therefore we copied them into the table as they are.
Commonly, scattering papers are missing an indication which QED effects were taken into account.
Additionally, sometimes the analysis have been ``fine-tuned'' to improve the goodness of the fit and/or resulting uncertainties. E.g.,~in Ref.~\cite{Kessler1975Muonic}, all transitions containing $3d$ states have been excluded from the analysis because of the anomaly in $\Delta 3d$ fine structure, which largely impacts the value of~$\chi^2$.
However, we believe it is useful to have a complete list of historically reported data for an overview and to highlight the complexity of the rms radius evaluation.

\end{document}